\documentstyle[preprint,eqsecnum,aps]{revtex}
\def\mtc#1#2#3{\multicolumn{#1}{#2}{#3}}

\def\ie{{\it i.e.\ }}

\def\ql{quasi-local}

\def\cc{cosmic censorship}
\newcommand{\spt}{spacetime}
\def\spl{spacelike}
\def\tl{timelike}
\def\sing{singularity}
\def\sings{singularities}

\begin{document}
\draft
\preprint{TIT/HEP-246/COSMO-41}
\title{Causal feature of central singularity and gravitational mass}
\author{Tatsuhiko Koike\thanks{JSPS fellow,
        e-mail address: bartok@phys.titech.ac.jp},
        Hisashi Onozawa\thanks{
        e-mail address: onozawa@phys.titech.ac.jp}
        and Masaru Siino\thanks{JSPS fellow,
        e-mail address: msiino@phys.titech.ac.jp}}
\address{Department of Physics, Tokyo Institute of Technology,
Oh-Okayama, Meguro-ku, Tokyo 152, Japan}
\date{December, 1993}
\maketitle
\begin{abstract}
Mass of singularity is defined,
 and its relation to whether the singularity is spacelike,
 timelike or null is discussed for spherically symmetric
 spacetimes.
 It is shown that if the mass of singularity is positive
 (negative) the singularity is non-timelike (non-spacelike).
 The connection between the sign of the mass and the force
on a particle is also discussed.
\end{abstract}

\section{Introduction}
\label{introduction}
 In discussing \cc~\cite{Pen69} of singular \spt s
in general relativity
it is important whether the singularity is ``spacelike'', ``timelike'',
or ``null'', which can be defined by conformal embedding.
 A \spl\ \sing\ immediately implies existence of an event
 horizon and save the \cc.
On the contrary, a \tl\ \sing\ is naked, hence violates the \cc.

 It is important
 to know what physical quantity determines
the causal feature of a \sing.
This quantity should be defined locally
since the causal feature is a local concept.

 The concept of \ql\ mass has been proposed by many authors
\cite{Kom59,Haw68,Pen82,LuVi82,KCD88,DoMa90,Hay93-1}.
Though asymptotic mass is well-defined such as the ADM~\cite{ADM60} or
the Bondi~\cite{BVM62} mass,
there is no satisfactory definitions of \ql\ mass.
When we discuss an appropriate definition of \ql\ mass,
its physical implication is no less important than mathematical
property to be considered as energy.
If we regard the \ql\ mass as the generalization of the
Schwarzschild mass, we may expect in a general spacetime that
the \ql\ mass play similar roles as the Schwarzschild mass.

 Let us consider the singularity $r=0$ of Schwarzschild spacetime,
\begin{equation}
  ds^2=-(1-2mr^{-1})dt^2+(1-2mr^{-1})^{-1}dr^2
        +r^2(d\theta^2+\sin\theta d\phi^2).
\end{equation}
 The usual definition of \ql\ mass
$M(t,r)$ in spherical symmetric spacetimes gives
$M(t,r)=m$.
 The mass of the singularity can be defined to be the limit of $M(t,r)$
when $r\rightarrow0$, which is $m$ in this case.
 The singularity is spacelike in the
case $m>0$ and is timelike in the case $m<0$.

In this paper we define mass of singularity
and discuss the relation between the mass of singularity and
whether it is spacelike, timelike or null.
We concentrated on spherically symmetric spacetimes because
they are essentially two-dimensional and
have a conformal embedding to the flat two-dimensional flat space
with infinity and singularity being its boundary.
We investigate the statement in general spherically symmetric:
if the mass of the singularity is positive, the singularity is spacelike,
and if it is negative, the singularity is timelike.
Further if we see that the negative mass implies the existence
of naked singularities, it may be possible to prove the cosmic
censorship theorem by discussing the inadequateness of negative
mass.

In \S 2, we define the terms in the above statement precisely.
Whether it holds or not is investigated in \S 3.
The \S 4 reveals the motion of a particle near the negative mass singularity.
In negative mass Schwarzschild spacetime no particle hits the singularity.
The final section is devoted to conclusions and discussions.

\section{Mass of singularity in spherically symmetric spacetimes}
We assume that the spacetime manifold $(M,g_{ab})$
is an $S^2$-bundle over a two-dimensional manifold $(M_1,(g_1)_{ab})$
so that the metric is given as
\begin{equation}
(g)_{ab}=(g_1)_{ab}+(g_2)_{ab},
\end{equation}
where $(g_2)_{ab}$ is the metric of a sphere
characterized by the area $A=4\pi R^2$ with $R$ being a (positive)
function on $M_1$.

Two-dimensional space $M_1$,
or at least a part of it under consideration,
is conformally embedded into
two-dimensional flat space
$(\tilde M_1,(\tilde g_1)_{ab})=(\tilde M_1,\eta_{ab})$ as
\begin{eqnarray}
& &\tilde M_1=M_1\cup \partial M,\\
& &(\tilde g_1)_{ab}=\eta_{ab}=\Omega^2 (g_1)_{ab}.
\end{eqnarray}
 Let us call the boundary on which $R=0$ central.
 We consider point $p$ of the central boundary
and the neighborhood $U$ of $p$ in $\tilde M_1$.
 The function $R$ is continuous on $M$ and the central boundary.
 We assume that the function $R$ and either $\Omega$ or $\Omega^{-1}$
can be continuously extended to
$U\cap\partial M_1$.

The line element is written in coordinates as
\begin{eqnarray}
ds^2& &=ds^2_1+ds^2_2,
\nonumber\\
ds_1^2& &=-2\Omega^{-2}(u,v)dudv=-a^2(t,r)dt^2+b^2(t,r)dr^2,
\nonumber\\
ds_2^2& &=R^2(d\theta^2+\sin^2\theta d\phi^2).
\label{eq:metric}
\end{eqnarray}
Dual-null tetrad can be chosen as
\begin{eqnarray}
& &l^a=\Omega(\partial_u)^a
      ={1\over\sqrt2\,ab}(b(\partial_t)^a+a(\partial_r)^a),\nonumber\\
& &n^a=\Omega(\partial_v)^a
      ={1\over\sqrt2\,ab}(b(\partial_t)^a-a(\partial_r)^a),\nonumber\\
& &(e_1)^a=R^{-1}(\partial_\theta)^a,\nonumber\\
& &(e_2)^a=R^{-1}(\sin\theta)^{-1}(\partial_\phi)^a;\nonumber\\
\end{eqnarray}
where $l^a$ and $n^a$ are the out-going and in-going null vectors,
respectively.

 In spherically symmetric spacetimes the known \ql\ masses such as
the Penrose mass and the Hawking mass fall into the standard one
$R^3(\Phi_{11}+\Lambda-\Phi_2)$.
The Hawking's expression for $S^2(q)$, the round sphere determined by
$q\in M_1$ is given by
\begin{eqnarray}
M(q)&=&{1\over 32\pi^{3/2}}A^{1/2}\int_S\mu(\cal R+\theta_+\theta_-)\nonumber\\
    &=& {R^3\over4}(\cal R+\theta_+\theta_-)
  \label{eq:M(q)}
\end{eqnarray}
where $\mu$ is the volume form on $M_2$.
The scalar curvature $\cal R$ of $S^2(q)$ and
the expansions $\theta_+$ and $\theta_-$ of out-going and in-going
null rays, respectively, are given
in our coordinates as
\begin{eqnarray}
& &{\cal R}=2R^{-2},\nonumber\\
& &\theta_+=2\Omega R^{-1}R_v
           =(1/\sqrt{2})(Rab)^{-1}(b R_t-a R_r),\nonumber\\
& &\theta_-=2\Omega R^{-1}R_u
           =(1/\sqrt{2})(Rab)^{-1}(b R_t+a R_r).
\label{eq:quantities}
\end{eqnarray}
 Eq. (\ref{eq:M(q)}) yields
\begin{eqnarray}
M(q)&=&R(\Omega^2 R_u R_v+{1\over2})\nonumber\\
    &=&{R\over2}\,(a^{-2} R_t^2-b^{-2} R_r^2+1).
\label{eq:M(q)coord}
\end{eqnarray}

 The mass $M_S$ of a point $p$
of the central boundary is defined to be the limit of $M(q)$
for $q\rightarrow p$, where $q$ is a point in $U\cap M$.

 From(\ref{eq:M(q)coord}) we have
\begin{equation}
M_S(p)=\lim_{q\rightarrow p}\Omega^2 R\, R_u R_v.
\label{eq:M_S}
\end{equation}

 We present some examples of spherically symmetric spacetimes
 with central singularity.
 The mass is calculated by the formula (\ref{eq:M(q)coord}).
The conformal diagram of the first two examples
are well known\cite{HaEl73}.
 The conformal diagram of the Tolman-Bondi \spt~\cite{Tol34,Bon47}
 is found by Eardley and Smarr~\cite{EaSm79}.
 The statement given in the bottom of \S \ref{introduction} is true
for all of them.

\begin{table}
\begin{tabular}{llcc}
\mtc{2}{c}{Spacetime}
                   & Causal feature
                             & Mass of central boundary  \\
\hline
\mtc{2}{l}{The Schwarzschild \spt}
                   &
                             &\\
$a^2=b^{-2}=1-2m r^{-1},R=r$
          &(positive-mass)
                   & \spl    & $m(>0)$  \\
\cline{2-4}
          & (negative-mass)
                   & \tl     & $m(<0)$  \\
\hline
\mtc{2}{l}{The Reissner-Nordstr\"{o}m \spt}
                   &         & \\
$a^2=b^{-2}=1-2m r^{-1}+e^2r^{-2},R=r$
          &        &\tl      & $-\infty$  \\
\hline
\mtc{2}{l}{The Friedmann-Robertson-Walker \spt}
                   &         &\\
$a=1, b=b(t),R=b c$,
          &        &         & \\
$b_t^2={8\over3}\pi b^2\rho(t)-k,$
          &        &         & \\
$c=\sin r,r$ or $\sinh r$ for $k=0,1,-1,$
          &(dust-filled)
                   &\spl     & $(4\pi/3)R^3\rho$(=constant$>0$)\\
\cline{2-4}
$\rho(t)$: density
          &(radiation-filled)
                   & \spl    & $+\infty$  \\
\hline
\mtc{2}{l}{The Tolman-Bondi \spt}
                   &         & \\
$a=1,R_r=W(r) b$,
          &        &         & \\
${1\over2}R_t^2-m_0(r)R^{-1}={1\over2}(W^2(r)-1)$,
          &        &         & \\
$m_0(r)=4\pi\int_0^r\rho_0 r^2dr$,
          &        &\spl     & $m_0(r)(>0)$ \\
\cline{3-4}
$W(r)$: any function
          &        & null    & 0  \\
\end{tabular}
        \caption{}
        \protect\label{tab:examples}
\end{table}

\section{Causal feature}
 If the boundary point $R=0$ is not singular the mass must be zero.
 So if the central mass is nonzero the boundary point at which
$R=0$ must be a singularity.
In this section we investigate causal feature of central
singularities.

\subsection{Trappedness}
The mass $M_S$ of the singularity $p$ immediately gives the informations
 of whether the points near the singularity are trapped or not.
{}From (\ref{eq:M(q)}) and (\ref{eq:quantities}) one has
\begin{equation}
 \theta_+\theta_-={2\over R^2}\left({2M(q)\over R}-1\right).
\end{equation}
If $M_S(p)>0$ then $\theta_+ \theta_-$ becomes positive
because $M(q)\rightarrow M_S(p)>0$ and $R\rightarrow0$
as $q$ approaches $p$.
This means that the points near the singularity are trapped.
If $M_S(p)<0$ then $\theta_+ \theta_-$ becomes negative
because $M(q)\rightarrow M_S(p)<0$.
This means that the points near the singularity are not trapped.
Therefore, one cannot see positive-mass singularities.

\subsection{Spacelikeness and timelikeness of singularity}
The argument in the previous section suggests the following.
In the positive-mass case both of the expansions
$\theta_+ \theta_-$ are negative
(positive), which implies that both of the future
(past) directed out-going and in-going null geodesics
approaches the $R=0$ singularity.
In the negative-mass case one of them approaches and the other
goes away from the singularity.
One may intuitively think that positive-mass singularities are
spacelike and negative-mass singularities are timelike.
We discuss this more precisely.

 We define the boundary $R=0$ to be spacelike, timelike, or null
by the conformally related metric $(\tilde g_1)_{ab}=\eta_{ab}$.
 There must be a strictly increasing smooth function $F$ of $R$
such that $(dF)_a$ can be extended smoothly to the central boundary
and does not vanish there.
Let us say that the central boundary is spacelike, timelike, or null
if $(\tilde g_1)^{ab} (dR)_a (dR)_b$ is positive, negative,
or zero, respectively.
 The norm is given by
\begin{equation}
  (\tilde g_1)^{ab} (dF)_a (dF)_b=-2F_{,R}{}^2R_u R_v.
  \label{eq:norm}
\end{equation}

 By (\ref{eq:M_S}), if $R \Omega^2F_{,R}{}^{-2}$ is bounded,
positive mass implies
that the norm (\ref{eq:norm}) is negative infinity
so that the singularities must be spacelike.
 If $R \Omega^2F_{,R}{}^{-2}$
 is not bounded the singular boundary is \tl\ or null.
 An example of the positive mass with a null singular boundary is
given by $R=-uv,\Omega=m^{1/2}R^{-1} (m>0)$,
 where the mass on a boundary point is given by $M_S=m$.
 Negative central mass implies that the singularity is timelike or null
by the same argument.

The zero-mass boundary point has the following possibilities:
it is a spacelike singularity, it is a timelike singularity,
it is a null singularity, or it is not a singularity.

\section{Dynamics of a particle near singularity}
In the negative-mass Schwarzschild spacetime any inertial
 observer cannot fall into the singularity,
though any observer can see it.
We expect that the situation is the same in general cases of
negative-mass \sing.

In the Schwarzschild case it is verified by writing down
the geodesic equations and  obtain the first integral of them.
 The equation can be treated as a
 problem of a particle in a potential which is infinite at
 $R=0$.
Since general spherically symmetric spacetimes do not admit
 calculating such first integrals explicitly we investigate whether
 the negative-mass singularity has infinite repulsive force
 by calculating the acceleration of an observer.

 If the singularity is timelike,
we can choose the coordinate $r$ to be the area radius $R$
in the metric
(\ref{eq:metric}):
\begin{equation}
  ds^2=-a^2(t,R)dt^2+b^2(t,R)dR^2+R^2(d\theta^2+\sin^2\theta d\phi^2).
\end{equation}
The formula (\ref{eq:M(q)coord}) implies
 that $b$ is expressed by the \ql\  mass as
\begin{equation}
  b^2=\left(1-{2M(t,R)\over R}\right)^{-1}.
\end{equation}
Let $u^a$ be the 4-velocity of an observer with constant $R$
 and $s^a$ be a radial unit vector normal to $u^a$ \ie
\begin{equation}
 u^a=a^{-1}(\partial_t)^a,\;\;  s^a=b^{-1}(\partial_R)^a.
\end{equation}
The acceleration of this observer is
\begin{eqnarray}
  f&=&s_a u^b\nabla_b u^a \nonumber \\
   &=&{a_{,R}\over a}\left(1-{2M(t,R)\over R}\right)^{1\over2}
\end{eqnarray}
where $\nabla_a$ is the covariant derivative of $g_{ab}$.
The $(tt)$- and $(RR)$-components of Einstein's equation reduce to
\begin{eqnarray}
 M_{,R}&=&4\pi R^2\rho,
\label{eq:ein(tt)}
\\
f&=&{1\over R^2}\left(M(t,R)+4\pi R^3 p_1\right)
\left(1-{2M(t,R)\over R}\right)^{-{1\over2}},
\label{eq:ein(RR)}
\end{eqnarray}
where $\rho=T_{ab}u^au^b$ and $p_1=T_{ab}s^as^b$
with $T_{ab}$ being the energy-momentum tensor.
Equation (\ref{eq:ein(tt)}) corresponds to the Gauss law
\begin{equation}
M(t,R)=M_S+4\pi\int_0^R R^2\rho dR.
  \label{eq:gauss law}
\end{equation}
Equation (\ref{eq:ein(tt)}) is the expression of the
acceleration of a particle caused by gravitational potential.
The second term in the first parenthesis is a general
relativistic effect.
Using this equation, the acceleration near the central
singularity is given by
\begin{equation}
  f\rightarrow\lim_{q\rightarrow p}{|M_S|^{1/2}\over R^{3/2}}
\left(-1+{4\pi R^3 p_1\over |M_S|}\right).
\end{equation}
 The observer needs an infinite amount of inward acceleration
if $p_1$ behaves as ${\cal O}(R^{-3+\epsilon})\;(\epsilon>0)$
or it is nonpositive to stay on the constant $R$ world-line.
This means that the observer feels infinite repulsive force near
the singularity.

Of course, (\ref{eq:ein(RR)}) is valid when the surface
does not contain \sings
Since the repulsive force disperses positive-mass particles the
negative-mass region causes instability.
Large amount of positive pressure could prevent
such instabilities formalism of naked singularities.

\section{Conclusions and discussions}
We showed that in spherically symmetric spacetimes
positive-mass \sings\ cannot be naked.
If the mass of the central singularity is positive the
singularity is spacelike or null, and
if negative it is timelike or null.
Negative \ql\ mass causes as repulsive force on a point particle
in the same way as Newtonian gravity.
Whereas positive pressure weaken the repulsive force as a general
relativistic effect.
Sufficient positive pressure keeps from negative-mass \sings\
from appearing in the \spt.

There was no null singularities with mass $M = 0$
among well-known exact solutions we have presented in \S 2.
We might expect that a certain condition forbid them.

Since the sign of singularity mass determines the causal feature
of the singularity,
we may translate the cosmic censorship conjecture into a problem
of singularity mass.
If conditions imposed on the matter restricts the sign of
singularity mass to be positive
spacetimes with naked singularity are ruled out.

In the case of spacetimes without spherical symmetry, it is
difficult to investigate causal feature of singularities,
since general spacetimes do not admit a conformal embedding into
a compact manifold with boundary.
In such a spacetime the various definitions of \ql\ mass give
different values.
Our arguments suggest that the Hawking mass is preferred
to the other \ql\ masses for the definition of mass of singularity.

\section*{Acknowledgments}
We are grateful to Ken-ichi Nakao and Kouji Nakamura for discussions.
We thank Professor Akio Hosoya for continuous encouraging.
The research is supported by the Scientific Research Fund of the
Ministry of Education, Science and Culture (T. K. and M. S.).

\end{document}